\documentclass[manuscript]{aastexmod}

\usepackage{graphicx}
\def\gsim{\;\rlap{\lower 2.5pt \hbox{$\sim$}}\raise 1.5pt\hbox{$>$}\;}
\def\lsim{\;\rlap{\lower 2.5pt  \hbox{$\sim$}}\raise 1.5pt\hbox{$<$}\;}
\newcommand\beq{\begin{equation}}
\newcommand\eeq{\end{equation}}

\usepackage{fancyhdr}
\begin{document}

\section*{}
\vspace{-0.5in}

\Large
\centerline{\bf Disruption of a Proto-Planetary Disk by the}
\centerline{\bf Black Hole at the Milky Way Centre}

\normalsize
\medskip
\centerline{Ruth A. Murray-Clay\footnote{rmurray-clay@cfa.harvard.edu} \& Abraham Loeb}
\medskip

\centerline{\it Institute for Theory \& Computation}
\centerline{\it Harvard-Smithsonian Center for Astrophysics}
\centerline{\it 60 Garden Street, Cambridge, MA 02138, USA}

\vskip 0.2in
\hrule
\vskip 0.2in

\nobreak

{\bf Recently, an ionized cloud of gas was discovered plunging toward
the supermassive black hole, SgrA*, at the centre of the Milky
Way. The cloud is being tidally disrupted along its path to closest
approach at $\sim$3100 Schwarzschild radii from the black hole. Here,
we show that the observed properties of this cloud of gas can
naturally be produced by a proto-planetary disk surrounding a low-mass
star, which was scattered from the observed ring of young stars
orbiting SgrA*.  As the young star approaches the black hole, its
disk experiences both photo-evaporation and tidal disruption,
producing a cloud.  Our model implies that planets form in the
Galactic centre, and that tidal debris from proto-planetary disks can
flag low mass stars which are otherwise too faint to be detected.}

\vskip 0.2in
\hrule
\vskip 0.2in

\noindent
{\bf Introduction}

Observations of the galactic centre recently yielded a cloud of ionized 
gas and dust falling inward on a nearly radial orbit.\cite{f1}
The cloud will reach pericenter in the summer of 2013, approaching 
our galaxy's central black hole, SgrA*, at a distance of only 270 AU.  
As this plunge progresses, tidal gravity from the black hole will disrupt 
the infalling cloud, providing a unique probe of gas flow near SgrA*.

The apocenter of the cloud's orbit, at $r_{\rm apo} = 8400$ AU = 0.04
pc from the black hole, coincides with the inner edge of the ring of
young stars orbiting SgrA*, and the plane of the cloud's orbit
coincides with that of the ring.\cite{f1,f2} The ring's
age, estimated from its population of O/WR stars, is $\sim$4--8
Myr.\cite{f3} At ages $\lesssim$3 Myr, most low mass stars host
proto-planetary gas disks\cite{f4} with radii of order 100
AU,\cite{f5} and $\sim$$1/5$ of stars with mass $0.1-1 M_\odot$ retain
their disks at ages of $\sim$5 Myr.\cite{f6,f7} Given the
black hole mass of $M_{\rm BH} = 4.3\times 10^6
M_\odot$,\cite{f8,f9} the tidal radius around a star of
mass $m_\star$ at a distance $r=0.04~{\rm pc}$ from SgrA* is $d_t \sim
r(m_\star/3M_{\rm BH})^{1/3} \sim 40{\rm AU}(m_\star/M_\odot)^{1/3}$.
A solar mass star could therefore host a stable disk with a radius of
$d_{\rm out} \sim d_t/3 \sim 12$ AU on a roughly circular orbit near
the inner edge of the young stellar ring.  Similarly, an M-dwarf with
mass $m_* = 0.3 M_\odot$ could host a stable disk having radius
$d_{\rm out} \sim 8$ AU.

We suggest that the newly discovered gas cloud\cite{f1} surrounds such
a star, which was scattered away from its original ring orbit and is
currently plunging toward the supermassive black hole (Figure 1).  The
star itself is too low mass to be observable, but the debris produced
through the disruption of its proto-planetary disk allowed it to be
detected.  We first calculate the properties of the system at the
cloud's observed location and argue that they match the
observations---an ionized cloud $\sim$100 AU in radius with density
$n\sim 3\times10^5$ cm$^{-3}$, an electron temperature of $10^4$K, and
a dust temperature of $\sim$550K, trailed by a stream of gas.  We then
provide predictions regarding the evolution of the cloud as it
approaches pericenter.  Finally, we demonstrate that the probability
of producing such an object is plausibly high, and we calculate the
implied rate of mass deposition by this process within the young
stellar ring.

\noindent
{\bf Results}

\noindent
{\bf Mass loss.} Although the tidal radius for a solar-mass star at
$r=0.04$pc is 40AU, the tidal radius at the cloud's pericenter
distance of $r_p = 270$AU $= 10^{-3}$pc is only $d_t = 1$AU
$(m_*/M_\odot)^{1/3}$.  At the most recently observed epoch,\cite{f1}
the cloud was approximately $6r_p$ from the black hole, with a tidal
radius of 6AU $(m_*/M_\odot)^{1/3}$.  Hence, the circumstellar disk is
already experiencing substantial tidal disruption.  At the same time,
the Galactic centre hosts an extreme flux of ionizing and Far
Ultraviolet (FUV) photons.  Proto-planetary disks in the ionizing
environment near O stars in the Trapezium cluster are known to
experience photoevaporation.\cite{f10} The stars experience mass loss
due to heating both by FUV and by Lyman limit photons.\cite{f11} The
former heat the disk to $\sim$$10^3$K, generating outflows at the
sound speed of $\sim$$3~{\rm km~s^{-1}}$, corresponding to the escape
velocity at a distance $\sim$100 AU $(m_*/M_\odot)$ from the star.
Well within this distance, the FUV-driven outflow is diminished,
though not entirely quenched.\cite{f12} At the $\sim$10 AU and smaller
distances of interest here and given the extreme ionizing environment,
Lyman continuum (ionizing) photons dominate the outflow, generating a
$\sim$$10^4$K ionized outflow moving at the sound speed of
$\sim$$10~{\rm km~s^{-1}}$.  This speed matches the escape velocity at
a distance of $d_{\rm esc} \sim 10$ AU $(m_*/M_\odot)$ from the star.
Loss from smaller distances occurs at a reduced rate, but still
generates a $\sim$$10~{\rm km~s^{-1}}$ outflow by the time the gas
reaches $d_{\rm esc}$.

Which process---tidal stripping or photoevaporation---dominates mass
loss from the disk?  Currently, tidal stripping dominates the
unbinding of mass from the star, and at large distances from the star
tidal stripping determines the ultimate fate of the gas.  However, the
outflow properties of the observed cloud are nevertheless currently
determined by photoevaporation.  This can be understood as follows.
Gas at a distance $d>d_t$ from its host star is accelerated by the tidal
potential to a relative speed of $\Delta v$ as it moves of order its
own radius away from its host star: $\Delta v \sim (GM_{\rm
BH}/r^2)(d/r)(d/\Delta v)$, so that $\Delta v \sim v_{\rm
circ}(d/r)$, with $v_{\rm circ} = (GM_{\rm BH}/r)^{1/2}$.  For $d = d_t$, $\Delta v \sim v_{\rm circ}
(m_*/M_{\rm BH})^{1/3}$.
At the cloud's current separation from SgrA*, $r\approx 6\times 10^{-3}~{\rm
pc} \approx 1300~{\rm AU}$, $v_{\rm circ} \approx 1700~{\rm km~s^{-1}}$
and $\Delta v \sim 10~{\rm km~s^{-1}} (m_*/M_\odot)^{1/3}$ at the tidal radius, comparable to the wind outflow rate.  Hence, the motion of tidally decoupled gas is dominated by the tidal field.  
At earlier times in the star's plunge, $\Delta v$ was smaller,
meaning that wind gas flowed out of the tidal radius faster than tidally disrupted gas.  Currently, near the disk edge, the dynamics of previously ejected gas are set by properties of the wind, while on the $\sim$100 AU scale of the cloud, tidal evolution overwhelms wind motions.

Although the current mass disruption rate of the proto-planetary disk,
$\dot M_{\rm dis}$, is larger than the wind outflow rate, $\dot
M_{w}$, wind gas emitted at earlier times dominates the currently
observed cloud.  In the short time that the infalling star has spent
in an enhanced tidal field with $d_t < d_{\rm out}$, the disk has only had
the opportunity to expand by a fraction of its $\sim$10 AU size.  At
$d = 8$AU, the time since $d_t = d$ along the infalling star's orbit
is $\Delta t = 3$yr for $m_* = 0.3M_\odot$.  Decoupled material has
traveled only $\sim$$\Delta v\Delta t \sim 4$ AU further from the host
star in that time. Figure 2 illustrates this point.  We ask how far a
test particle, released at a given disk radius, $d$, from the star
when $d_t = d$ and moving only under the gravity of the black hole,
will be from its current orbit.  Exterior to the decoupled material,
gas originally launched in a wind dominates.

The tidal decoupling rate in the disk is $\dot M_{\rm dis} \sim
2\pi\Sigma d_t \dot d_t \sim 2\pi\Sigma d_t \dot r (m_\star/3M_{\rm
BH})^{1/3}$, where $\Sigma(d)$ is the original surface density of the
disk and $\dot r \approx 2000~{\rm km~s^{-1}}$ at the current position
in the cloud's orbit.  For illustration, we choose a profile similar
to the minimum-mass solar nebula: $\Sigma = \Sigma_0 (d/d_0)^{-1}$
with $\Sigma_0 = 2\times 10^3~{\rm g~cm^2}$ and $d_0 = 1$AU.\cite{f5}
This choice yields a current $\dot M_{\rm dis} \sim 3\times 10^{-3}
M_\odot~{\rm yr}^{-1} (m_*/M_\odot)^{1/3}$, where we set $d$ equal to
the current tidal radius.  Photoevaporation, on the other hand, gives
mass loss rates\cite{f13,f14} of $\dot M_{w} \sim
3\times10^{-10} (d_{\rm out}/10~{\rm AU})^{3/2}
(\Phi_{i,49}/D_{\rm pc}^2)^{1/2} M_\odot $ yr$^{-1}$ for disks with sizes
$d_{\rm out} > d_{\rm esc}$, where $\Phi_{i,49}$ is the ionizing luminosity,
$\Phi_i$, in units of $10^{49}$ s$^{-1}$ of a source at distance $D$
from the disk, with $D_{\rm pc} = D/(1{\rm pc})$.  

The photoevaporation mass loss can be derived as follows.  At $d=
d_{\rm esc}$, the escape velocity is comparable to the wind's sound
speed $c_s = 10$km s$^{-1}$.  For $d \gtrsim d_{\rm esc}$, $\dot M_{w}
\approx 4\pi d^2 m_p n_b c_s$, where $m_p$ is the mass of a proton and
we have neglected an order unity correction arising from the
non-sphericity of the flow.  We refer to the surface layer of the disk
within which photoionization heats disk gas to $\sim$$10^4$K as the
``base of the wind."  At the base of the wind, a balance between
photoionization and radiative recombination yields a number density of
$n_b \sim (\Phi_i/4\pi D^2)^{1/2} ( \alpha_{\rm rec}d)^{-1/2}$, where
$\alpha_{\rm rec} = 2.6 \times 10^{-13}$ cm$^3$ s$^{-1}$ is the Case B
radiative recombination coefficient for hydrogen at a temperature of
$10^4$ K, and we have used the fact that the optical depth to ionizing
photons is unity.  Hence, $\dot M \sim (4\pi/\alpha_{\rm rec})^{1/2} m_p
c_s d^{3/2} (\Phi_i/D^2)^{1/2}$.  Though this expression is only good
to order of magnitude, its coefficient is in fortuitously good
agreement with more detailed wind models.\cite{f13,f14}
Observations yield $10^{50.8}$ s$^{-1}$ Lyman continuum photons in the
central parsec\cite{f1,f15}, corresponding to $\Phi_{i,49}
= 63$.  Using $D=1$pc, $(\Phi_{i,49}/D_{\rm pc}^2)^{1/2} = 8$.  The
central concentration of S-stars within 0.01 pc, which we estimate to
contribute $\Phi_{i,49} = 0.2$ from each of ten approximately
10$M_\odot$ stars comparable to the second-most luminous Trapezium
star,\cite{f13} yields a small total of $\Phi_{i,49} = 2$ but in a
more concentrated region.  At the current position of the cloud, these
stars contribute $(\Phi_{i,49}/D_{\rm pc}^2)^{1/2} \sim 230$ for $D =
6\times 10^{-3}$pc.  At $D = 0.04$pc, this number is 35.  For $d =
10$AU and smaller, mass loss from these ionizing fluxes dominates over
FUV-driven mass loss.  Using an intermediate value of
$(\Phi_{i,49}/D_{\rm pc}^2)^{1/2} = 100$, $\dot M_{w} \sim 3\times10^{-8}
M_\odot~{\rm yr^{-1}} (d_{\rm out}/10{\rm AU})^{3/2}$.  On the cloud's
original orbit, $\dot M_{w} \sim 10^{-8} M_\odot~{\rm yr^{-1}} (d_{\rm
out}/10{\rm AU})^{3/2}$, allowing our nominal disk, which contains
$\sim$$10^{-2} M_\odot$ between 1 and 8 AU, to survive for
$\sim$$10^6$ yr.  Disk masses several times larger are plausible, and
hence a proto-planetary disk could have survived until the current
time on the star's birth orbit in the ring.

\noindent
{\bf Dynamics of stripped gas.} Currently, gas farther than $\sim$12
AU from the star (for $m_* = 0.3 M_\odot$; see Figure 2) was
originally ejected in the photoevaporative wind.  This ejected
material (which starts in a ring-like configuration) itself undergoes
tidal stripping.  Along the star's original orbit, the extent of the
wind moving at $\sim 10~{\rm km~s^{-1}}$ is set by the original 24 AU
tidal radius.  Since gas requires only 10 years to travel 24 AU at
$10~{\rm km~s^{-1}}$, this wind scale applies even if the wind region
was disrupted by close stellar encounters or more distant encounters
with the black hole at some point in the past.

As the star plunges toward the Galactic centre, its disk and wind are
pulled off in shells as the tidal radius shrinks (Figure 3).  The time
for a parcel of wind to travel at $10~{\rm km~s^{-1}}$ from 10 to 100
AU is comparable to the 70 year time to plunge to pericenter on the
cloud's current orbit.  This wind-generated cloud in turn experiences
tidal disruption.  By its current location, the original wind cloud
will have reached an extent of a few hundred AU.  Figure 2 illustrates
this extent for $m_* = 0.3 M_\odot$.  We note that although previous
close encounters with the black hole may have stripped the disk to
smaller than its original size, the disk wind is regenerated over each
orbit.  As long as the disk size exceeds $d_{\rm esc}$ and most of the
disk mass remains intact, our wind calculation remains valid.  If the
disk is stripped to smaller sizes, a wind will still be blown, but
with reduced $\dot M_w$.  

Ram pressure of the ambient gas exceeds the ram pressure of the
photoevaporative wind when $n_{amb}v_\star^2 > n v_w^2$, where
$n_{amb}$ is the ambient number density of gas and $v_\star$ is the
star's velocity along its orbit.  The characteristic density of
ambient gas within the central 1.5pc is
$n_{amb}\sim10^3$cm$^{-3}$.\cite{f20} Models place the density at
$\sim 3\times 10^2$--$6\times 10^3~{\rm cm^{-3}}$ at the cloud's
current location and $\sim (1$--$5)\times 10^2~{\rm cm^{-3}}$ on its
original orbit.\cite{f16} Along the star's original orbit, $v_\star
\approx 700~{\rm km~s^{-1}}$, and the ram pressure force from the disk
wind roughly balances ram pressure with the ambient medium at the
star's tidal radius.  Currently, $v_\star \approx 2300~{\rm
km~s^{-1}}$, so ram pressure with the surrounding medium has increased
by 1--2 orders of magnitude at comparable separations from the star,
and the tidally-disrupted photoevaporative wind is undergoing ram
pressure stripping.  Nevertheless, at the current tidal radius, the
two pressures remain in rough balance and our estimates of the mass
loss rate are therefore valid.  As the cloud continues its plunge
toward the super-massive black hole, its outer (tidally detached)
extent will be shaped by ram pressure stripping.

Figure 4a displays the inferred ionized density of the cloud as
a function of radial scale.  From the total Br$\gamma$ line
luminosity, the discovery paper \cite{f1} calculates an electron
density of $2.6\times10^5 (d/125 AU)^{-3/2}$, in excellent agreement
with our prediction.  In Figure 4a, we also plot the contribution to
the total line luminosity as a function of radial scale.  Since the
luminosity is proportional to $n^2 d^3$, this contribution peaks at
the outer edge of the disk, but the contribution from the extended
cloud falls off slowly, as $1/d$, so that about 1/5 of the line
luminosity comes from 100 AU scales.  We note that though the majority
of Br$\gamma$ emission comes from the 10--20 AU scale of the tidally
expanded disk, the majority of the mass in the cloud is at large
scales since the cloud mass is proportional to $n d^3 \propto d$.  At
these large scales, full hydrodynamic simulations including ram
pressure stripping and tidal gravity in 3D are required to match in
detail the surface brightness, shape, and velocity width of the
observed emission.  One might naively expect the surface of the cloud
to be Kelvin-Helmholtz unstable.\cite{f1} However, observations of
cold fronts in X-ray clusters indicate that gas clouds moving at a
Mach number of order unity through a hot ($\sim$ keV) ambient medium
maintain a smooth surface, probably due to "magnetic
draping".\cite{f17}

\noindent
{\bf Discussion} 

We predict (Figure 4b) that the total Br$\gamma$
luminosity of the cloud will increase as it approaches pericenter.
This future evolution of a debris cloud around a low-mass star on a
Keplerian orbit is easily distinguishable from that of a
pressure-confined cloud with no self gravity or central mass supply.
We further predict that with better resolution, the specific intensity
of the line should increase since most of the emission is coming from
a smaller spread in velocities than is currently resolved.
 
The dust in the wind, in analogy to dust in HII regions, does not
reach temperature equilibrium with the $10^4$K gas.  Gillessen et
al.~argue that the dust continuum emission at about 550 K comes from
small, transiently heated grains, having a total mass equal to
$\sim$$10^{-5}$ the total gas mass of the cloud.\cite{f1} Additional
colder dust may be present.  This relatively small dust mass may
result from grain growth and settling in the proto-planetary disk,
leaving relatively few small grains available to be lofted into the
photoevaporative wind.  The neutral, cooler disk does not contribute
substantially to the observed dust emission.
  
As demonstrated above, a $\sim 3\times10^{-2}M_\odot$ disk can survive
on the star's original orbit for $\sim$3 Myr.  In the Supplementary
Discussion, we discuss the likely rate at which such disks could be
scattered onto orbits comparable to the observed gas cloud.  This rate
primarily depends on the highest mass scatterers in the ring of young
stars around SgrA*. It could increase considerably above a minimum
value $0.1\%$ if the stellar mass function extends above the observed
$60M_\odot$ limit (as observed elsewhere in the Galaxy), if there are
intermediate mass black holes \cite{f18} of $10^2$--$10^3M_\odot$, or
if compact star clusters (such as the observed clusters IRS 13N or IRS
13E) act as massive scatterers.  Finally, if a 0.03 $M_\odot$ disk is
disrupted by the black hole every $\sim$$5\times 10^5$yr, the related
events deliver $\sim 6\times 10^{-8}M_\odot~{\rm yr}^{-1}$ to the
inner 0.01 pc around SgrA*, comparable to the presently inferred
accretion rate onto the black hole.\cite{f19}

We note that alternative explanations invoking mass loss during rare
evolutionary phases of stars (such as a proto-planetary nebula around
an emerging white dwarf or a debris envelope around a merged blue
straggler), have a much lower probability of occurance than our
scenario.  They share the requirement for a rare kick, but involve
progenitors which are much less abundant than low-mass stars in the
ring of young stars around SgrA*.

The existence of proto-planetary disks in galactic nuclei has
important implications: it could lead to a fragmentation cascade to
comets, asteroids, and dust around quasars,\cite{f20} and to bright
flares due to the tidal disruption of planets,\cite{f21} as well as
transits of hypervelocity stars.\cite{f22} In the Milky Way centre,
the debris of proto-planetary disks offers a new probe of the low mass
end of the mass distribution of stars which are too faint to be
detected otherwise.

\normalsize
\vskip 0.2in
\noindent
{\bf Acknowledgements.} We thank Andi Burkert, Reinhard Genzel, and Chris
McKee for stimulating discussions.  AL was supported in part by NSF
grant AST-0907890 and NASA grants NNX08AL43G and NNA09DB30A.

\vskip 0.2in
\noindent
{\bf Author contributions.} Both authors originated the idea for the
project and worked out collaboratively its general details.
R.M.C. performed the N-body simulation described in the Supplementary
Discussion.

\vskip 0.2in
\noindent
{\bf Competing Financial Interests:} The authors have no competing
financial interests.

\vskip 1in

\begin{figure*}[hptb]
\begin{center}
\includegraphics{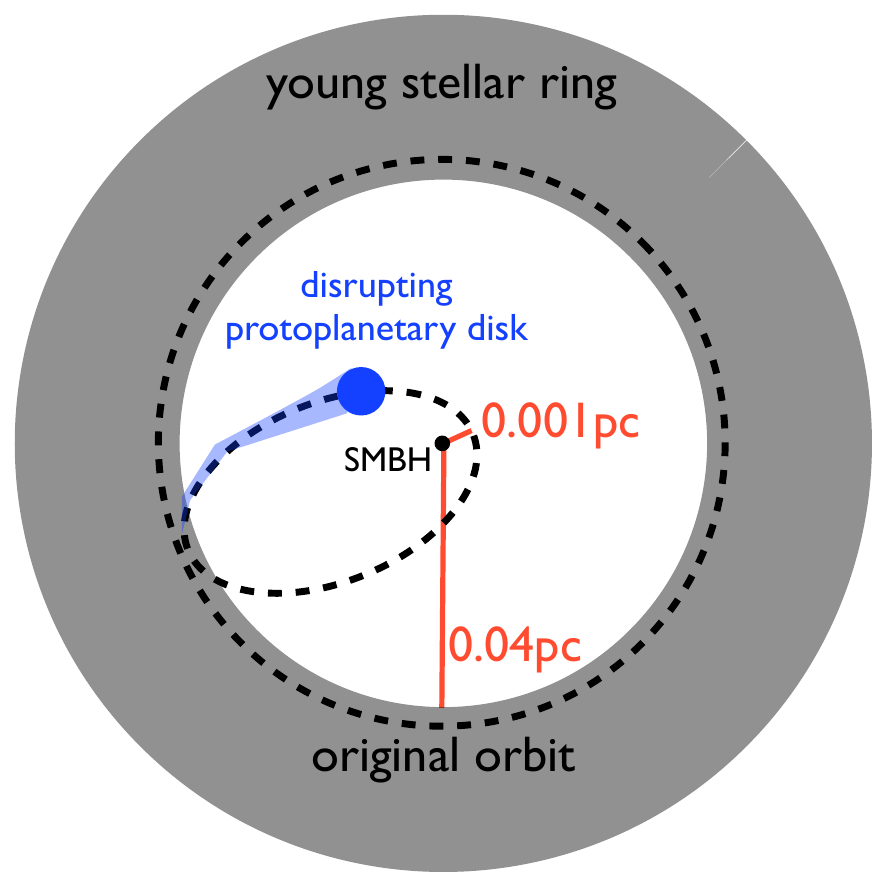}
\end{center}
\caption{{\bf Proto-planetary disk plunging toward the galactic centre.}  Schematic illustration of our scenario.  A young, low-mass
star hosting a proto-planetary disk is dynamically dislodged from its original orbit (dashed circle) in the
Galactic centre's young stellar ring (gray).  As it plunges toward the
supermassive black hole (SMBH) at the Galactic centre (orbit indicated by dashed ellipse), the disk is
photoevaporated and tidally disrupted, generating an extended dusty
cloud around the star (blue circle).  A trail of dust and gas is deposited along the star's orbit (extended blue trail).}
\end{figure*}

\begin{figure*}[hptb]
\begin{center}
\includegraphics{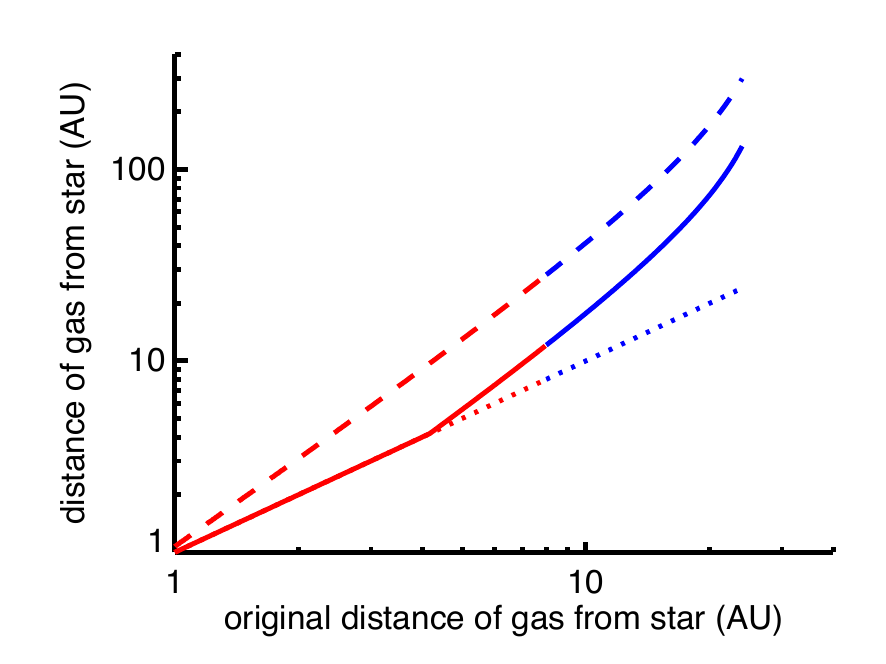}
\end{center}
\caption{\small {\bf Expansion of the infalling gas.} As the star plunges toward the Galactic centre, its
disk and the surrounding photoevaporative wind are tidally disrupted.
For a fiducial stellar mass of $m_* = 0.3M_\odot$, we plot the
distance, $d$, of gas from the host star at the current epoch (solid
line) and at the time of pericenter passage (dashed line) as a
function of separation from the host star at the time of tidal
decoupling.  The dotted line represents no change.  We assume that the
material is instantaneously decoupled from the star when its
separation equals the tidal radius and that it subsequently moves as a
test particle in the gravitational field of the supermassive black
hole.  At the time of its decoupling, the test particle is started
with the orbital parameters that the cloud had at time $d/v_*$ prior
to decoupling, and the full separation between the test particle and
the star is thereafter calculated as a function of time.   
Red portions of the curve represent gas initially in the disk, while
blue portions represent gas initially in the photoevaporative cloud.
For gas currently observed in the cloud, the tidal evolution depicted
here dominates the wind structure at large scales, while the wind
outflow at $\sim 10~{\rm km~s^{-1}}$ dominates the structure of
recently ejected gas near the disk rim.  At pericenter, tidal
evolution determines the structure of the entire wind.}
\end{figure*}

\begin{figure*}[hptb]
\centerline{ $
\begin{array}{@{}c@{}c@{}}
\includegraphics{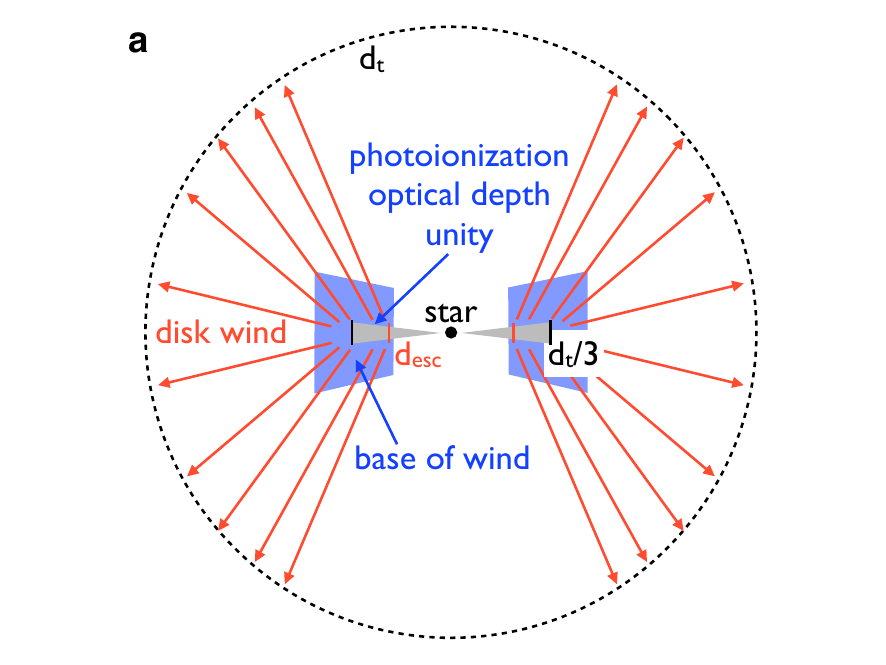} &
\includegraphics{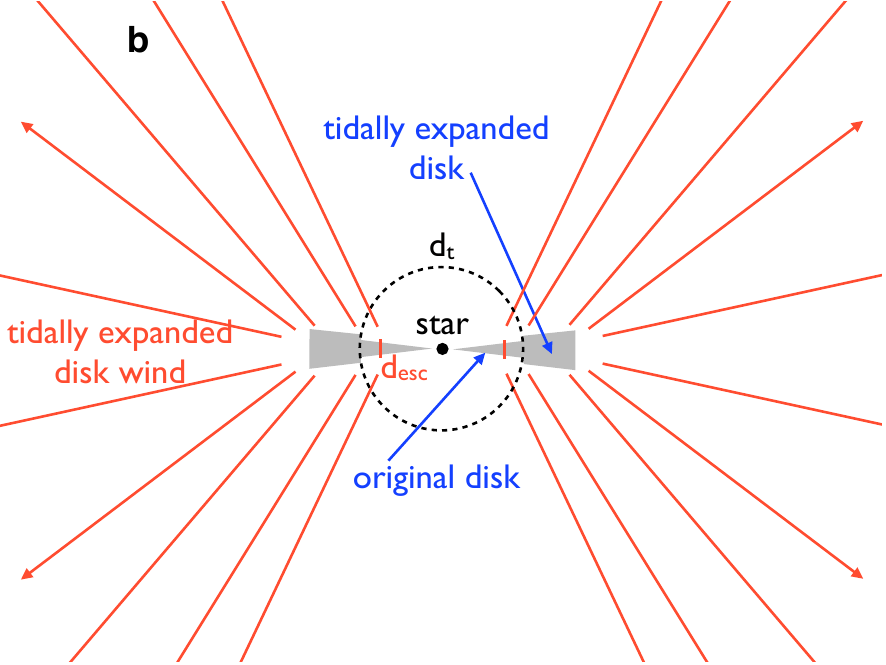}
\end{array} $
}
\caption{{\bf Disk and wind structure.} Schematic diagram of the disk and wind structure on the
star's original orbit in the young stellar ring ({\bf a}) and at the
current epoch ({\bf b}).  On its original orbit, the star (black circle) hosts a protoplanetary disk (gray), which is limited in extent (vertical black line) to approximately one third of the star's tidal radius, $d_t$ (dashed circle). Photoionization efficiently launches a wind beyond $d_{\rm esc}$, where gas is heated enough that its thermal velocity is comparable to the stellar escape velocity (vertical red line). From the wind launching region (blue), gas flows outward (red arrows) until it passes the tidal radius and is stripped from the star.  At the current epoch, $d_t$ has shrunk, causing both the disk and the wind to tidally expand.}
\end{figure*}

\begin{figure*}[hptb]
\centerline{ $
\begin{array}{@{}c@{}c@{}}
\includegraphics{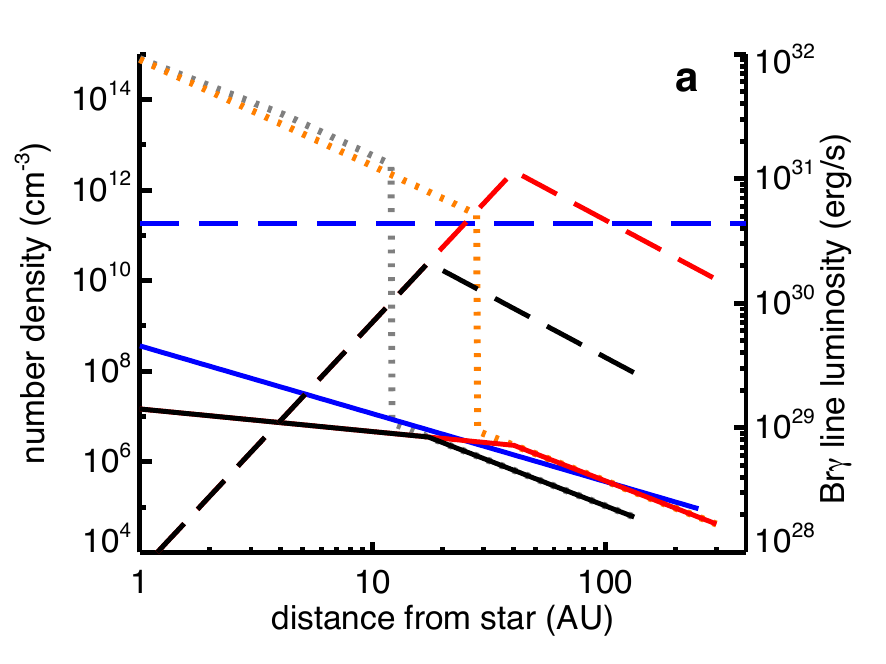} &
\includegraphics{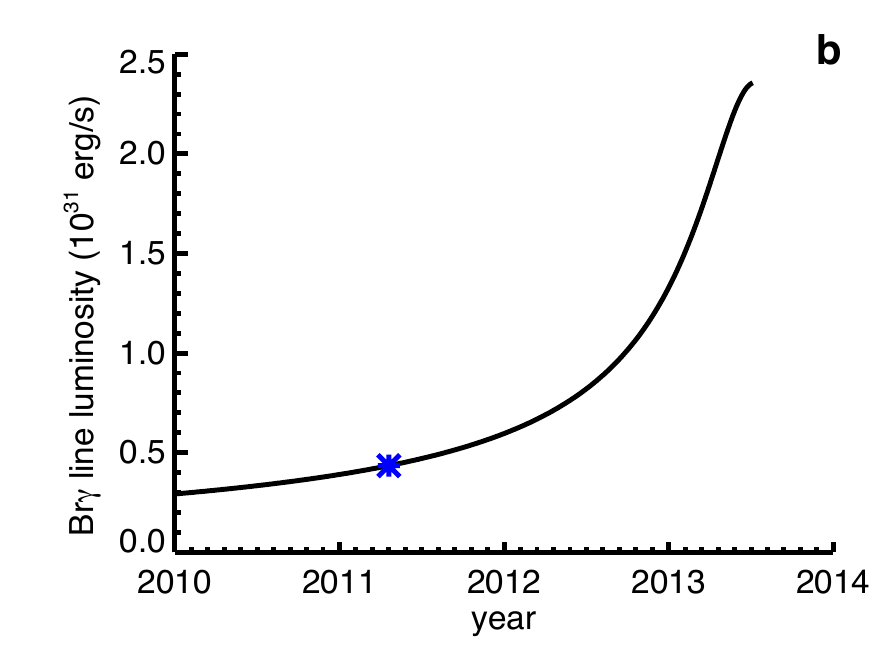}
\end{array} $
}
\caption{\footnotesize {\bf Observational predictions.}  Ionized gas in the disk wind generates Br$\gamma$ line
luminosities in good agreement with observations.  {\bf (a), left axis:} The
cloud's ionized gas number density is inferred from the observed
Br$\gamma$ line luminosity to be\cite{f1} $2.6\times 10^5 (d/125{\rm
AU})^{-3/2}$ cm$^{-3}$ for emission dominated by gas at a distance $d$
from the center of the cloud (blue, solid).  We approximate the
neutral number density in the disk as $n = m_p^{-1}
\Sigma_0(d_{i}/d_0)^{-1} h_d^{-1} (d_{i}/d)^2$, where $d_{i}$ is the
initial separation of the gas from the host star and we use a disk
scale-height $h_d \sim 0.1 d$.  We show the ionized density, $n_+$
at the current epoch (black, solid) and at the time of
pericenter passage (red, solid) for $m_* = 0.3 M_\odot$.  Where wind
gas dominates, we estimate $n_+ = \dot M_w(2\pi m_p d^2 v_w)^{-1}$
with $\dot M_w = 3\times10^{-8} (d_{\rm out}/10{\rm AU})^{3/2}
M_\odot/$yr, assuming that the wind geometry is not yet spherically
symmetric.  We choose $d_{\rm out}$ to be the current outer limit of
the tidally disrupting disk.  This estimate is not good at the largest
separations, where tidal evolution and ram pressure stripping modify
the gas density (see text).  Where the disk itself is present, the
neutral gas number density in the disk (gray dotted, current; orange
dotted, pericenter) is 5--8 orders of magnitude larger than the
ionized number density at the disk surface.
We therefore set $n_+ = n_b$, the number density at the base of the
ionized wind (see text).  Near the outer edge of the disk, the
observed number density is in excellent agreement with the modeled
density, given the approximate nature of this model. 
{\bf (a), right axis:} Br$\gamma$ luminosity as a function of distance, $d$, from the
infalling star, calculated using $L_{Br\gamma} = 2.35\times
10^{-27}n_+^2 (4/3) \pi d^3$ (in c.g.s. units) now (black, dashed) and
at pericenter (red, dashed) for the same model.  The currently
measured line luminosity is included for reference (blue, dashed),
and is in good agreement with the model.  Most of the total integrated
line emission is coming from near the current disk edge at 10--20 AU,
but $\sim$1/5 of the emission comes from $\sim$100 AU scales.  {\bf (b)} Total Br$\gamma$ line luminosity integrated over
our model disk plus wind as a function of time, calculated as $L_{Br\gamma} = 2.35\times
10^{-27} \times 2\pi \int n_+^2 d^2 \;{\rm d}d$ (in c.g.s. units; using $2\pi$ to adjust for a non-spherical wind).
We predict that the total line luminosity will increase by approximately a factor of 5 from its reported value (blue asterisk) as the star approaches pericenter in 2013.} 
\end{figure*}

\pagebreak

\setcounter{page}{1}

\begin{center}
{\bf \normalsize SUPPLEMENTARY INFORMATION \\
for ``Disruption of a Proto-Planetary Disk by the \\ Black Hole at the Milky Way Centre"}
\end{center}

\normalsize
\centerline{Ruth A. Murray-Clay \& Abraham Loeb}

\begin{suppfigure}[hptb]
\centerline{ $
\begin{array}{@{}c@{}c@{}}
\multicolumn{2}{c}{\mbox{\bf Supplementary Figures}} \vspace{0.5in} \\
\includegraphics[width=0.5\textwidth]{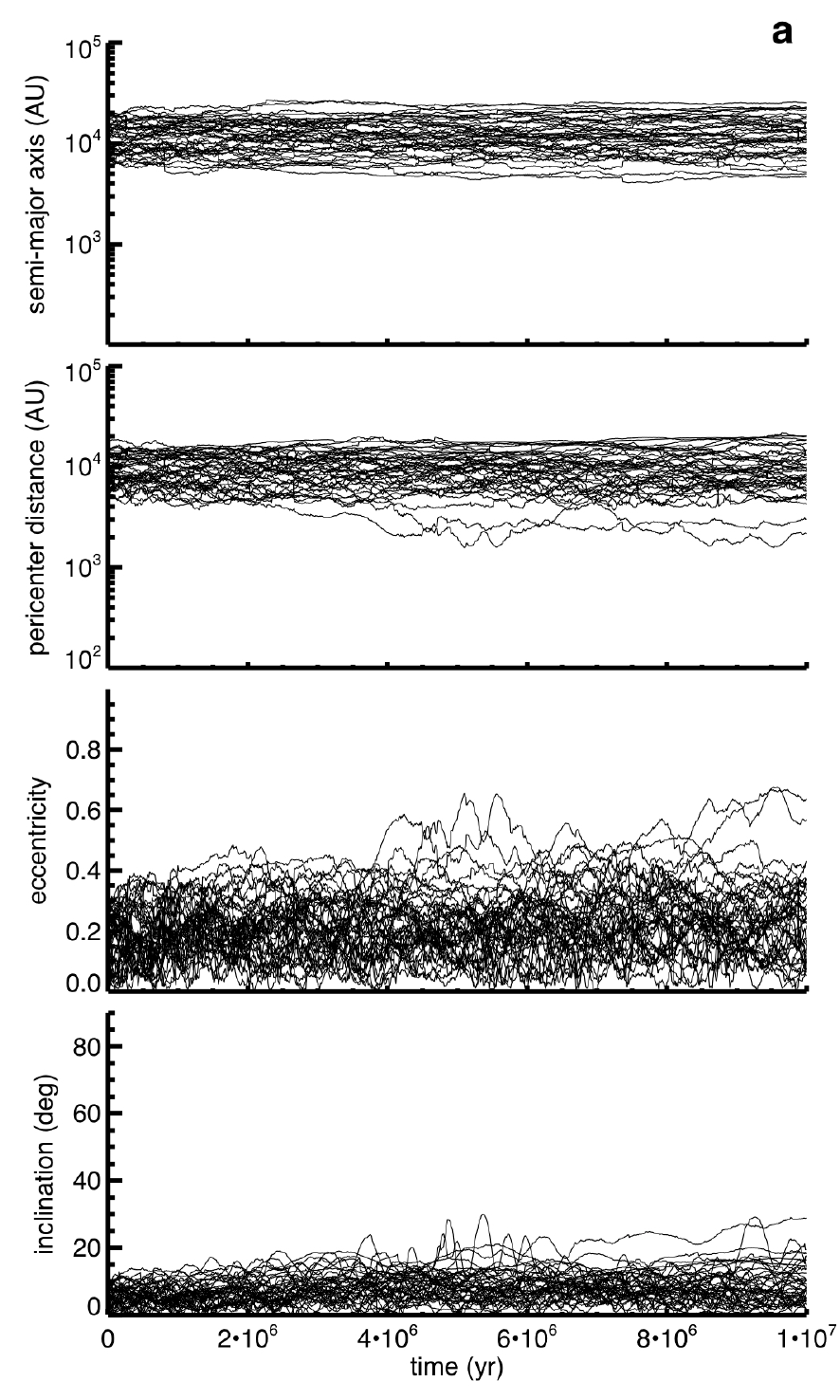} &
\includegraphics[width=0.5\textwidth]{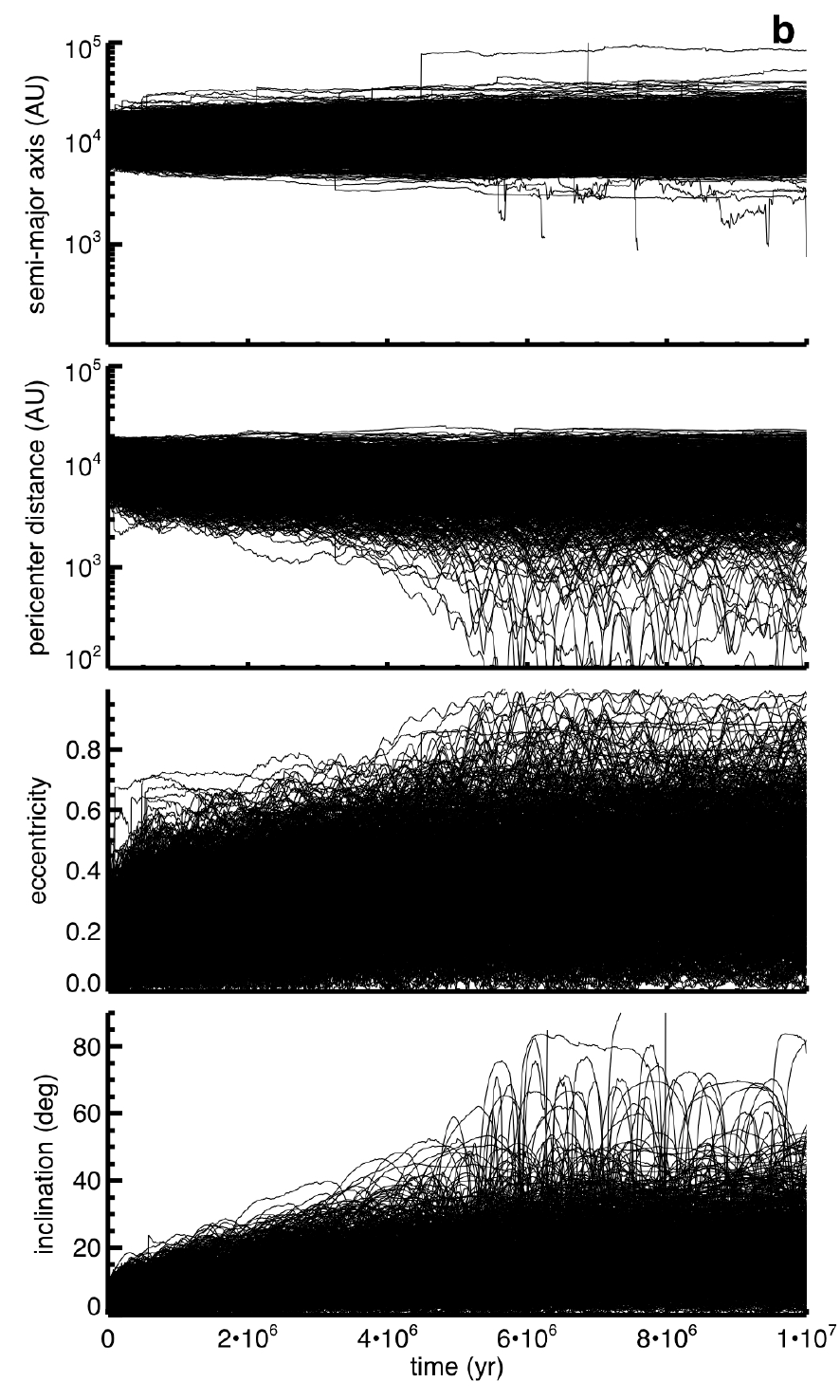}
\end{array} $
}
\caption{{\bf  Dynamical properties of stars.} Semi-major
axes, pericenter distances, eccentricities, and inclinations for
perturbing stars ($m_* = 60M_\odot$; {\bf a}) and low-mass stars (test
particles; {\bf b}) as functions of time for our first simulation,
described in the text.  The perturbers maintain properties consistent
with the young stellar ring, while a subset of low mass stars are
delivered onto low angular momentum orbits.}
\end{suppfigure}

\pagebreak

\begin{suppfigure}[hptb]
\vspace{1in}
\centerline{ $
\begin{array}{@{}c@{}c@{}}
\includegraphics[width=0.5\textwidth]{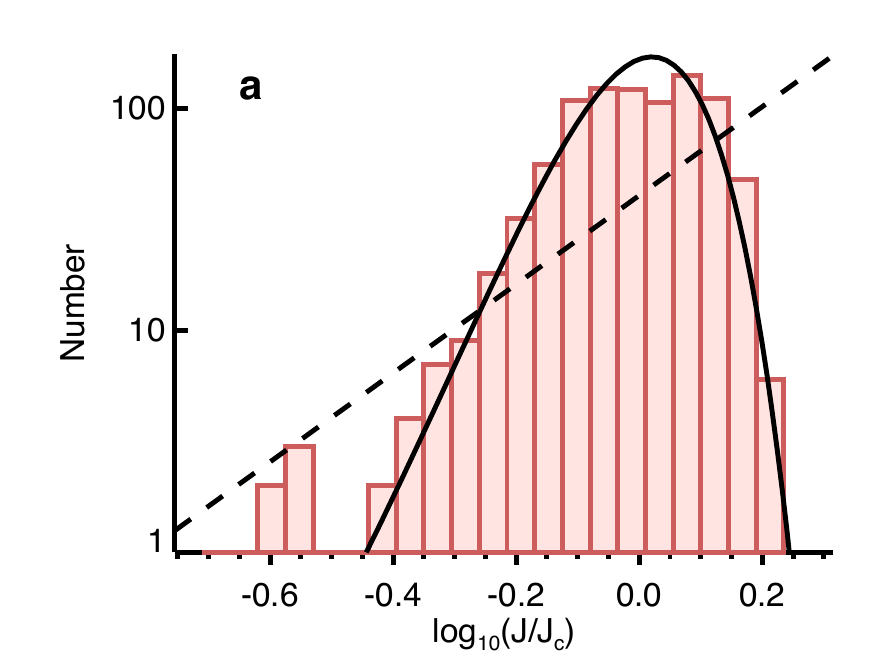} &
\includegraphics[width=0.5\textwidth]{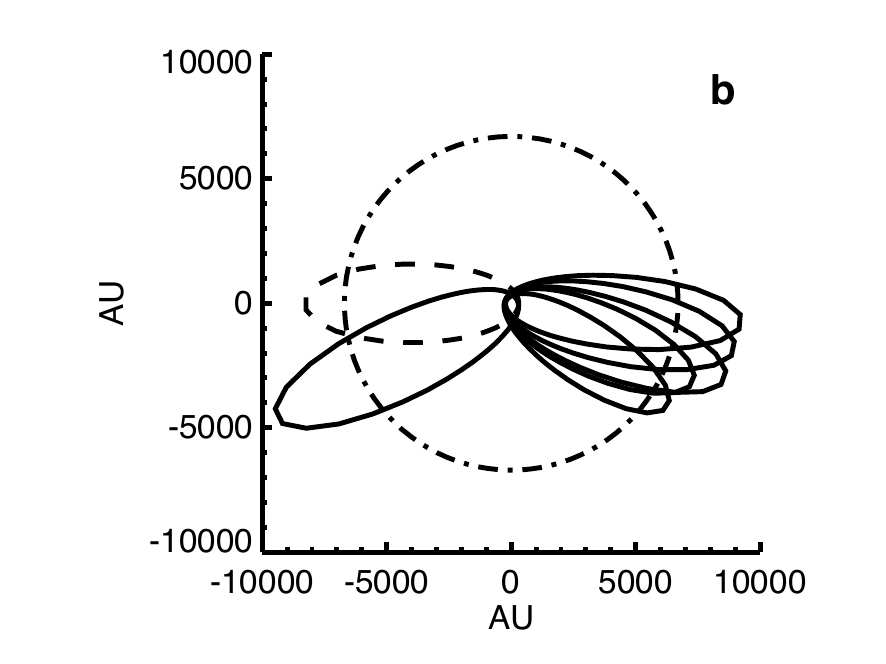}
\end{array} $
}
\caption{{\bf  Results from N-body
simulations.} {\bf (a)} Histogram of angular momenta of low mass
stars at time $6\times 10^6$ years for the same simulation shown in
Figure 5.  The number of stars is well matched by 1-D angular momentum
diffusion (equation S5; solid).  The expectation for a relaxed system,
$N = N_* (J/J_c)^2 \Delta$ (dashed), is plotted for reference.  For
the observed cloud, $\log_{10}(J/J_c) = -0.6$.  {\bf (b)} Orbits of
simulated stars (solid) that reach pericenter distances smaller than
that of the observed cloud (dashed) in less than $6\times 10^6$
years. The inner edge of the young stellar ring is plotted for
reference (dot dashed).}
\end{suppfigure}

\begin{suppfigure}[hptb]
\centerline{ $
\begin{array}{@{}c@{}c@{}}
\includegraphics[width=0.5\textwidth]{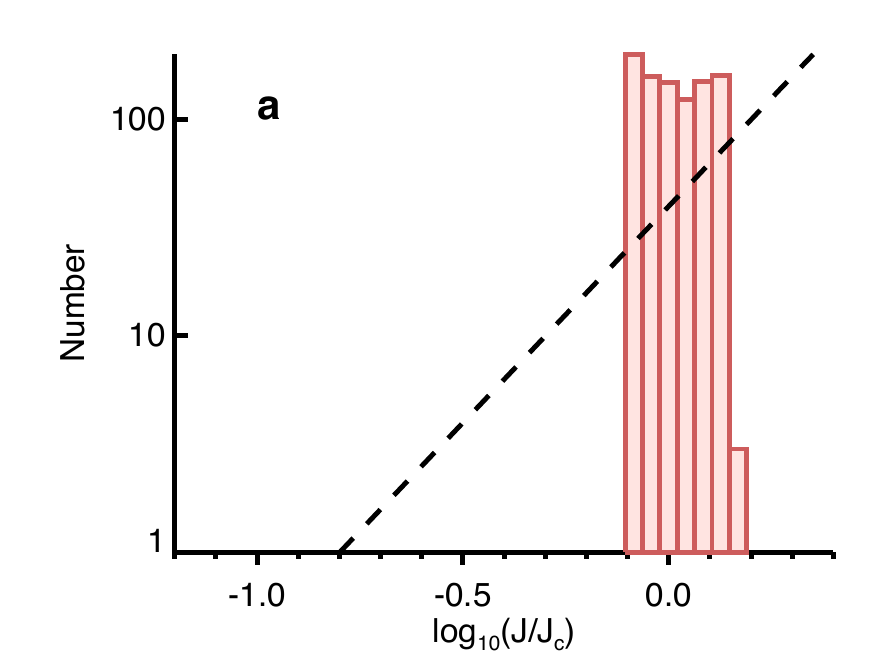} \includegraphics{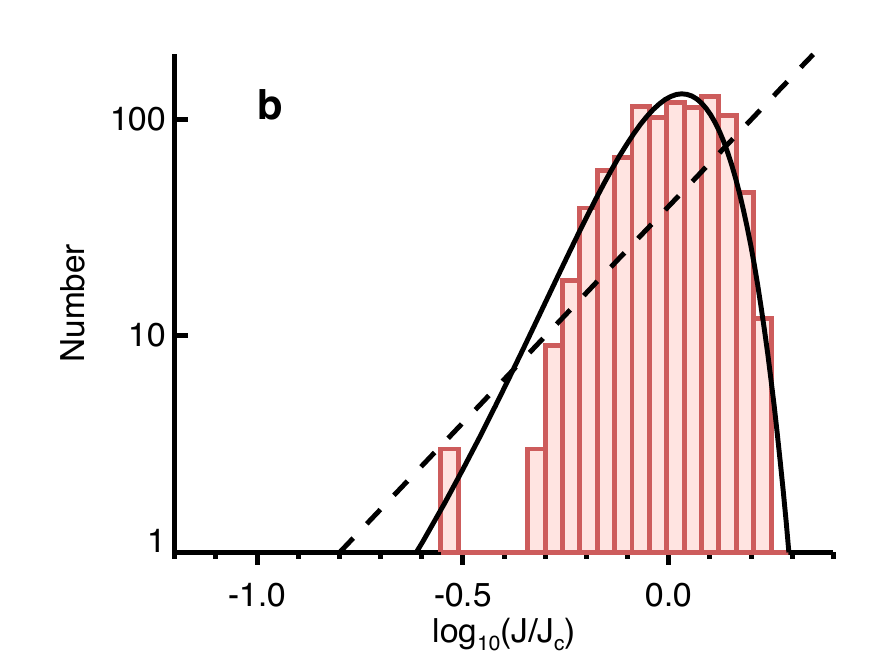} \\
\includegraphics[width=0.5\textwidth]{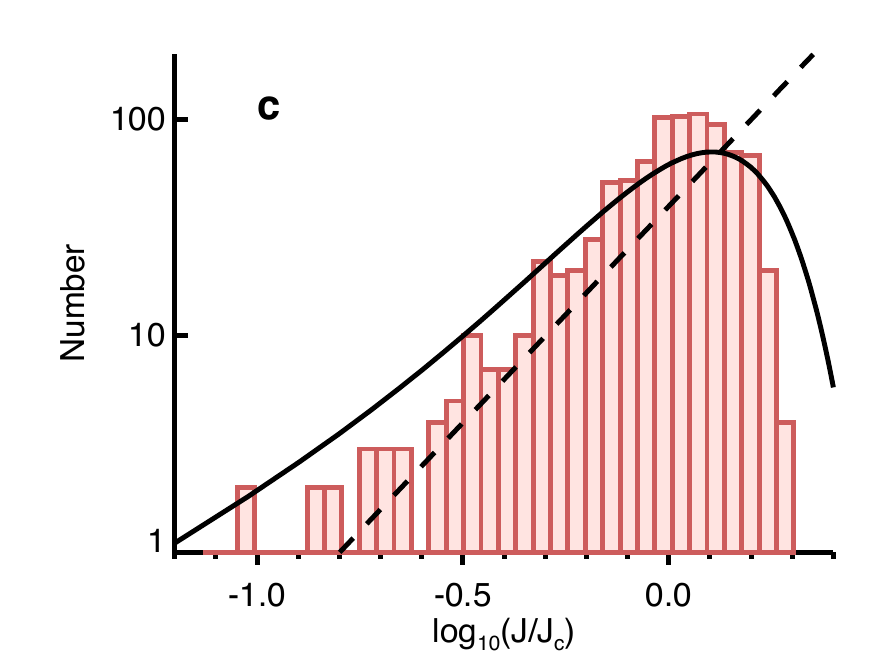} \includegraphics{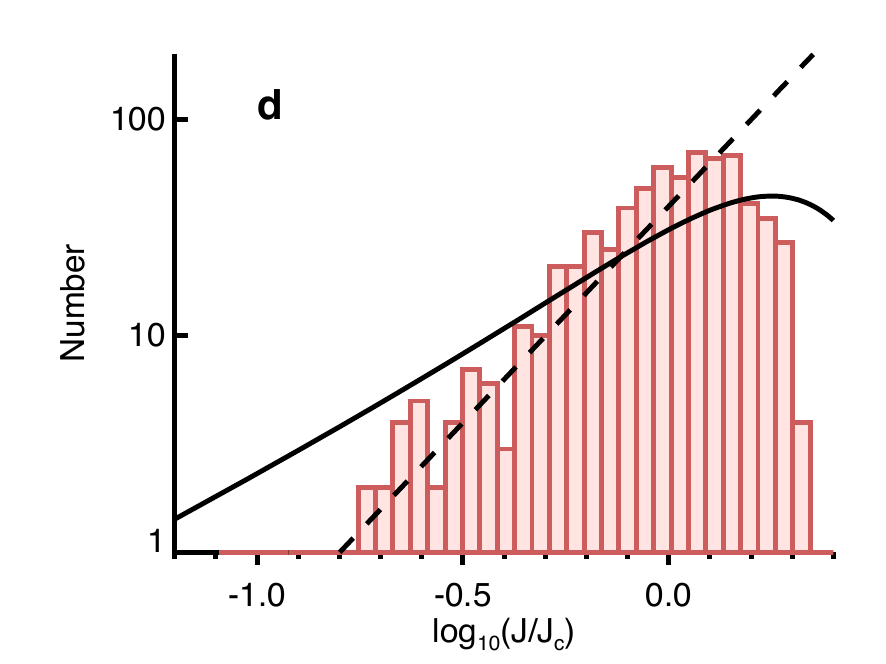}
\end{array} $
}
\caption{{\bf Angular momenta of low mass
stars.}  We show histograms at times $0$ ({\bf a}), $6\times 10^6$yr ({\bf b}), $2.5\times
10^7$yr ({\bf c}), and $10^8$yr ({\bf d}) in a simulation containing 24 perturbers, each
with $m_* = 100M_\odot$ (see text).  The low-mass stellar population,
which starts with evenly distributed angular momenta in the inner
ring, is first matched by the 1D diffusion equation (Eq. 5; solid) and
then by $N = N_* (J/J_c)^2 \Delta$ (dashed).}
\end{suppfigure}

\pagebreak

\section*{Supplementary Discussion}

\noindent
{\bf Scattering Probability.} Proto-planetary disks hosted by
young stars could be stripped by several possible mechanisms.  In the
main text, we have shown that on their initial orbits in the young
stellar ring, such disks are not depleted by mass loss in disk winds
over their several million year lifetimes.  However, the same
dynamical processes that could send a young star plunging toward the
Galactic centre could themselves lead to disk stripping.  In
particular, scattering events can strip disks and, if the star's
orbital evolution is too gradual, earlier encounters with the central
black hole could also produce stripping.  To arrive on its current
plunging orbit without losing its proto-planetary disk, a young star
must: {\it (i)} experience sufficient orbital evolution over its
lifetime due to scattering interactions at all distances to be placed
on a highly eccentric orbit while avoiding single encounters that
strip the disk, and then {\it (ii)} experience a final, strong
encounter that substantially alters the star's pericenter distance,
again without stripping its disk.  Requirement {\it (ii)} is needed so
that the disk's extent was not already truncated by tidal stripping at
a comparable pericenter location during a previous plunge toward the
black hole.

We note that disk winds can be driven by photoionization even if the
disk is truncated at closer distances to the host star than those
estimated above, so requirement {\it (ii)} is not strictly needed for
our model.  However, if requirement {\it (ii)} is not satisfied, the
mass loss rate in the disk's wind and hence the observed cloud density
will be lower than estimated in the main text.

In order to investigate these requirements, we first consider the
stripping potential of a single scattering encounter.  We demonstrate
that a final, strong encounter with a high mass star or intermediate
mass black hole \cite{f18} can change the pericenter distance of a
plunging star without stripping its proto-planetary disk.  We then show
that, given choices about the distribution of massive perturbers that
are consistent with observations, evolution of a low-mass star to a
highly eccentric orbit followed by a strong final encounter could
happen with reasonably high probability, all without disk stripping.

A single scattering encounter with a perturber of mass $m_{\rm pert}$
at impact parameter $b$ and with relative velocity at infinity $v_{\rm
enc}$ changes a star's velocity by $\Delta v_* \sim (Gm_{\rm
pert}/b^2)(2b/v_{\rm enc})$.  A given change in velocity $\Delta v_*$
hence requires an impact parameter $b \sim 2Gm_{\rm pert}/(v_{\rm
enc}\Delta v_*)$.  Such an encounter tidally strips disk gas at radii
$r_d$ where the induced relative velocity between the star and its
orbiting material, $\Delta v_{\rm tid} \sim (Gm_{\rm
pert}/b^2)(2r_d/b)(2b/v_{\rm enc})$ exceeds the star's escape velocity
$v_{\rm esc,*} =(2Gm_*/r_d)^{1/2}$.  Note that because we will be
interested in $\Delta v_* \leq v_{\rm enc}$, we can ignore
gravitational focusing in our estimate of $\Delta v_{\rm tid}$
(encounters are not parabolic, as is often assumed in calculations of
disk stripping during close stellar passages).  During an encounter
that changes the stellar velocity by $\Delta v_*$, the disk is
stripped at radii $r_d \gtrsim (b/2) (v_{\rm esc,*}/\Delta v_*)$, or
in other words, where $v_{\rm esc,*}^3 \lesssim 2(m_*/m_{\rm
pert})v_{\rm enc}(\Delta v_*)^2 $.  We choose to express the location
of stripping in terms of $v_{\rm esc,*}$ rather than $r_d$ because
disk winds are launched where the thermal velocity, $v_{\rm th}$, of
gas with temperature $T \sim$$10^4$K is comparable to the escape
velocity from the star.  The location, $r_d$, of this equality varies
as a function of stellar mass.  The change in stellar velocity
generated by an encounter that does {\it not} strip gas at the disk
radius with a given $v_{\rm esc,*}$ is limited to
\begin{eqnarray}
\Delta v_* \lesssim \left(\frac{m_{\rm pert}}{m_*}\frac{v_{\rm
esc,*}^3}{2v_{\rm enc}} \right)^{1/2}~~~~~~~~~~~~~~~~~~~~~~~~~~~~~(S1) \nonumber
\end{eqnarray}

We may now estimate the perturber mass required to change a plunging
star's pericenter distance by of order itself without truncating the
star's disk at the radius corresponding to $v_{\rm esc,*} = v_{\rm
th}$.  Since the star's orbit has an eccentricity of\cite{f1} $e =
0.938 \sim 1$, the specific angular momentum of its orbit around the
black hole is approximately $J \approx (2GM_{\rm BH} r_p)^{1/2}$,
where $r_p$ is the pericenter distance.  A change in the star's
orbital angular momentum of $\Delta J$ corresponds to a change in
pericenter distance $\Delta r_p \approx 2 r_p (\Delta J/J)$.  To
generate $\Delta r_p = r_p/2$, a single scattering event at apocenter
must change the star's apocenter velocity, $v_{\rm apo} = [GM_{\rm
BH}(1-e)/r_{\rm apo}]^{1/2}$, by $\Delta v_{\rm apo} \approx v_{\rm
apo}/4$.  The cloud's current, post-scattering, apocenter velocity of
$v_{\rm apo,f} = (3/4)v_{\rm apo} = 170~{\rm km~s^{-1}}$ requires
$\Delta v_* = \Delta v_{\rm apo} = v_{\rm apo,f}/3$.

A single scattering encounter near the apocenter of the star's orbit around the black hole can therefore generate $\Delta r_p = r_p/2$ without stripping the star's disk at the distance corresponding to $v_{\rm esc,*} = v_{\rm th}$ as long as the perturber has mass
\begin{eqnarray}\label{eqn-pertmasslim}
m_{\rm pert} &\gtrsim& \frac{2}{9}m_* \frac{v_{\rm enc} v_{\rm
apo,f}^2}{v_{\rm th}^3}~~~~~~~~~~~~~~~~~~~~~~~~~~~~~~~~~~~~~~~~~~(S2) \nonumber \\ &\sim& 100
M_\odot \left(\frac{m_*}{0.3 M_\odot}\right) \left(\frac{v_{\rm
enc}}{300 {\rm km~s^{-1}}}\right)\left(\frac{v_{\rm apo,f}}{170 {\rm
km~s^{-1}}}\right)^2\left(\frac{v_{\rm th}}{18 {\rm
km~s^{-1}}}\right)^{-3} \nonumber \\ &\sim& 700 M_\odot
\left(\frac{m_*}{1 M_\odot}\right) \left(\frac{v_{\rm enc}}{670 {\rm
km~s^{-1}}}\right)\left(\frac{v_{\rm apo,f}}{170 {\rm
km~s^{-1}}}\right)^2\left(\frac{v_{\rm th}}{18 {\rm
km~s^{-1}}}\right)^{-3} \nonumber \;\;.
\end{eqnarray}
In the numerical evaluation above, we have normalized $v_{\rm th}$ to
the isothermal sound speed for ionized hydrogen at $T = 2\times
10^4$K.  The exact ionized gas temperature depends on gas metallicity
(higher metallicity leads to more effective cooling) and on the
typical energies of ionizing photons (softer spectra generate lower
temperatures).  We choose two reference encounter velocities.  First,
$v_{\rm enc} = 300~{\rm km~s^{-1}}$ is the difference between
$(4/3)v_{\rm apo,f}$ and the velocity of a star also reaching
apocenter at $r_{\rm apo}$ but having an eccentricity of 0.4,
consistent with the average eccentricity of massive ring stars, which
is observed to be\cite{f15} $\sim 0.3$--$0.4$.  Near (but not exactly
at) both star's apocenters, a favorably oriented encounter could have
this velocity and reduce the angular momentum of the star by
approximately the amount estimated above.  Second, a perturber from an
isotropic population would typically have $v_{\rm enc} \sim (GM_{\rm
BH}/r_{\rm apo})^{1/2} = 670~{\rm km~s^{-1}}$.  From equation
(S2), a member of the stellar population of the
young ring could serve as the final perturber for a star with mass
$m_* \lesssim 0.3M_\odot$, given a favorable, but not implausible,
encounter geometry.  A solar mass star likely requires an encounter
with a more massive perturber, perhaps an intermediate mass black hole
with mass $M_{\rm IMBH} \sim 10^3M_\odot$ residing outside of the
young disk. Such black holes may be present in the Galactic centre,
and are not vilating existing dynamical constraints.\cite{f18} In
fact, Merritt et al.~(2009)\cite{f24} showed that the orbits of the
S-stars could be a natural consequence of the interaction between the
nuclear star cluster and an intermediate mass black hole (IMBH) with a
mass $\gtrsim 1.5\times 10^3M_\odot$ on a mildly eccentric orbit over
a period of a few Myr.  Again, equation (S2) is not
a strict requirement for our model, but if it is not satisfied, $\dot
M_w$ is reduced.

Having established that a final strong encounter can cause a
disk-laden star to plunge toward the black hole without stripping the
disk, we now consider the probability of such a history for a star in
the Galactic centre.  We calculate the rate of delivery of low mass
stars onto highly eccentric orbits, the probability of a final
encounter that substantially changes the star's pericenter distance,
and the likelihood that a star could be delivered onto a plunging
orbit without experiencing a disk-stripping encounter during its
lifetime.  We make choices about the mass function of stars in the
region that are simultaneously optimistic for our theory and
consistent with observations.

First, what is the rate of delivery of low-mass stars onto highly eccentric orbits?  
Over its lifetime $T$, each low-mass star is scattered multiple times
by ring stars of mass $m_{\rm pert}$ at impact parameters $b > b_{\rm
min}$, where $b_{\rm min}$ is the typical minimum encounter distance
after an elapsed time $T$.  The value of $b_{\rm min}$ is set by the
encounter rate $f_{\rm pert} (b) = [N_{\rm pert}/(\zeta\pi R^2 h)]\pi
b^2 v_{\rm rel}$, such that $f_{\rm pert} T= 1$ for $b = b_{\rm min}$.
To avoid disk disruption, $b_{\rm min}$ must be large enough that disk
disrupting encounters are not common.  We quantify this constraint
below.  Here, $N_{\rm pert}$ is the number of perturbing ring stars,
$v_{rel}$ is the typical relative velocity between perturbers and the
low-mass star, $R = r_{\rm apo} = 8400$AU is the local ring radius, $h
\sim \sigma_\star/\Omega$ is the ring scale height (with $h/R\sim
10^\circ$),\cite{f15} $\sigma_\star$ is the velocity dispersion of
ring stars, and $\Omega = (GM_{\rm BH}/R^3)^{1/2}$ is the angular
orbital velocity of the ring.  We include an order unity factor,
$\zeta$, to more accurately approximate the number density of
perturbers.  We will show below that $\zeta \sim 5$ given our fiducial
choices for $N_{\rm pert}$ and $R$.  The measured velocity dispersion
$\sigma_\star = h\Omega = 120~{\rm km~s^{-1}}$ corresponds to expected
eccentricities of $\sqrt{2}\sigma_\star/(R\Omega) = 0.25$, in rough
agreement with observed eccentricities of $e_* = 0.3-0.4$.  As a
compromise between these measured values, we define a stellar random
velocity $v_{\rm rand} = \sqrt{3}\sigma_\star \sim e_*R\Omega =
200~{\rm km~s^{-1}}$, and we set the typical relative rencounter
velocity $v_{\rm rel} = v_{\rm rand}$ during the low-mass star's early
orbital evolution, where we have plugged in $e_* = 0.3$.  Once the
star has reached a highly eccentric orbit, $v_{\rm rel} \sim \Omega R
- v_{\rm apo} > \sigma_\star$ near apocenter, and we adopt $v_{\rm
rel} = \Omega R$.

\noindent
{\bf Diffusion of angular momentum.} Transport of stars onto low
angular momentum orbits proceeds by angular momentum diffusion, as in
standard loss cone calculations.\cite{f25} At early times, the impact
parameter, $b_{\rm strong}$, required to generate a ``strong"
scattering that alters the star's random velocity by of order itself
is $b_{\rm strong} = Gm_{\rm pert}/v_{\rm rand}^2$, and each such
scattering generates a fractional angular momentum change of order
$v_{\rm rand}/(\Omega R)$.  More generally, as long as $b \ge b_{\rm
min} \ge b_{\rm strong}$, each star's specific angular momentum, $J
\approx Rv_{\rm apo}$, is altered by of order itself on a timescale
$T_{\rm diff} = J^2/D$, where the diffusion coefficient
\begin{eqnarray}\label{eqn-diffcoeff}
D &\sim& R^2 v_{\rm apo}^2 \frac{N_{\rm pert}}{\zeta \pi R^2 h}\pi \left(\frac{Gm_{\rm pert}}{v_{\rm apo}v_{\rm rel}}\right)^2 v_{\rm rel} \ln\Lambda ~~~~~~~~~~~~~~~~~~~~~~~~(S3) \nonumber
\\ 
&\sim& \frac{N_{\rm pert}}{\zeta h} \frac{G^2 m_{\rm pert}^2}{v_{\rm rel}} \ln\Lambda \nonumber \\
&\sim& J_c^2 \Omega {N_{\rm pert}\over \zeta} {\ln\Lambda} \left(\frac{m_{\rm pert}}{M_{\rm BH}}\right)^2 \frac{(\Omega R)^2}{\sigma_* v_{\rm rel}}\;\;, \nonumber
\end{eqnarray}
where $J_c = R^2\Omega$ is the specific angular momentum of a star on
a circular orbit at the ring radius $R$, and $\ln\Lambda\equiv \ln
(0.08~{\rm pc}/b_{\rm min})$ is the Coulomb logarithm due to multiple
gentle scatterings.  Though the diffusion time for stars with angular
momentum $J \ll J_c$ is shorter than the relaxation time $T_{\rm rel}
\equiv J_c^2/D$, by a factor\cite{f25} of $(J/J_c)^2$, the fraction of
phase space occupied by low angular momentum orbits is also
proportional to $(J/J_c)^2$, so that a population of $N_*$ low-mass
stars in the the young stellar ring produces low angular momentum
stars a rate\cite{f26}
\begin{eqnarray}\label{eqn-fstart}
f &\sim& \frac{N_*}{T_{\rm rel}} ~~~~~~~~~~~~~~~~~~~~~~~~~~~~~~~~~~~~~~~~~~(S4) \nonumber
 \\
&\sim& \Omega N_\star \frac{N_{\rm pert}}{\zeta} \ln\Lambda \left(\frac{m_{\rm pert}}{M_{\rm BH}}\right)^2 \frac{(\Omega R)^2}{\sigma_* v_{\rm rel}} \;\;, \nonumber
\end{eqnarray} 
as long as the system is sufficiently relaxed.  We return to the question of relaxation below.  At early times, $v_{\rm rel} \sim v_{\rm rand}$, and at late times $v_{\rm rel} \sim R\Omega$.  However, we note that at late times, the number of perturbers available with comparable relative velocities could easily be a factor of $R\Omega/v_{\rm rand} \sim 3$ higher given contributions from  stellar remnants in the isotropic population, so in estimating our rates, we employ $v_{\rm rel} = v_{\rm rand}$.  We note that though it plunges interior to the ring's extent as it approaches pericenter, the cloud nevertheless spends more than half of its orbital time within the ring (near apocenter).  As demonstrated above, disk survival requires $\Delta v_{\rm tid} \lesssim v_{\rm th}$, which translates to $b_{\rm min} \gtrsim [8 G^2m_{\rm pert}m_*/(v_{\rm enc}v_{\rm th}^3)]^{1/2}$. For $m_{\rm pert} = 60 M_\odot$, $v_{\rm enc} = 200~{\rm km~s^{-1}}$, and $m_* = 0.3 M_\odot$, disk survival requires $b_{\rm min} > 10$AU.  At the limit $b_{\rm min} = 10$ AU,  $\ln\Lambda \approx 7$.
Since $f \propto N_{\rm pert}m_{\rm pert}^2$, high-mass stars dominate the stellar orbital evolution.

\noindent
{\bf Mass function of scatterers.} For the top-heavy mass function of
stars inferred in the young stellar ring,\cite{f27} the number of
stars $N$ having mass $m$ follows $dN/dm \propto m^{-0.45\pm 0.3}$ in
the mass range $7M_\odot < m < 60M_\odot$, and the presence of
higher-mass stars is not excluded.  Most of the ring's mass is
associated with the highest mass perturbers, since $m^2dN/dm \propto
m^{1.55}$.  The inferred total mass in the young stellar ring
of\cite{f15,f27} $\sim 10^4 M_\odot$ could accommodate
$N\sim170$ perturbers with a mass $m_{pert}=60M_\odot$ each.  The
surface density of observed O/WR stars scales radially as\cite{f27}
$\propto$$r^{-1.4}$, extending from $0.8^" = 0.03$pc$=6700$AU to $12^"
=0.5$pc$=10^5$AU (using a distance to the Galactic centre of $R_0 =
8.33$kpc).  Approximately $(1/10)^{0.6} \approx 1/4$ of the ring's
O/WR stars reside at the inner scale of the ring near $r_{\rm apo} =
1^" \approx 0.04$pc, the location of the proposed initial stellar
orbit.  Hence, we adopt fiducial values of $m_{\rm pert} = 60 M_\odot$
and $N_{\rm pert} = 40$ in the vicinity of the cloud's $r_{\rm
apo}$. Integrating over the full disk (with surface density
$\Sigma_{\rm pert}(r)$), we find that $N_{\rm pert} = 40$ perturbers
lie between $R_1 = 6700$AU and $R_2 = 2\times 10^4$AU.  The volumetric
number density of perturbers at $R = 8400$AU is approximately
$\Sigma_{\rm pert}/(2h) = N_{\rm pert}/(\zeta \pi R^2 h)$ with $\zeta
= (20/3)(R_2^{0.6}-R_1^{0.6})/R^{0.6} = 5$.

In contrast, the mass function of observed stars both interior and
exterior to the young stellar ring is consistent with a
Salpeter/Kroupa initial mass function (IMF).\cite{f27} We suppose for
the sake of argument that the IMF in the young ring changes its slope
to the Salpeter value, $dN/dm \sim m^{-2.35}$ below a stellar mass
$m<m_{break}=7M_\odot$ and that this mass function extends down to
stars of $0.3M_\odot$, consistent with the observed turnover in the
IMF in other galactic environments \cite{f28} (but see also
\cite{f29}).  Given these assumptions, $N \propto m^{0.55}$ for $m >
m_{break}$ and $N \propto m^{-1.35}$ for $m < m_{break}$.  Then, the
inner ring would host $\sim$12 stars of $m=7M_\odot$, $N=170$ stars of
$m=1M_\odot$, and $N=900$ stars of $m=0.3 M_\odot$.  The presence of a
few thousand young low-mass stars in the ring is consistent with
limits from X-ray observations.\cite{f30} These choices generate
delivery rates of $f = 6\times10^{-7}$ yr$^{-1}$ for $m_* = M_\odot$
and $3\times 10^{-6}$ yr$^{-1}$ for $m_* = 0.3 M_\odot$.  Scattering
on binary stars further enhances our calculated rate.

\noindent
{\bf N-body simulations.} However, we must worry about the rate
calculation in equation (S4) because the young stellar ring is not
relaxed.\cite{f25} The angular momentum relaxation time $T_{\rm rel}
\sim 3\times 10^8$yr, and we are interested in timescales more than an
order of magnitude shorter than this.  We perform a set of N-body
simulations using the Mercury6 hybrid orbital integrator\cite{f31}
with a timestep of 10 years and and accuracy parameter of $10^{-12}$.
Supplementary Figures S1 and S2 show results from a simulation with 40 fully massive
perturbers, each with $m_* = 60 M_\odot$, and 900 test particles.
Both populations are distributed in semi-major axis from 6700AU to
20100AU with surface densities in semi-major axis proportional to
$r^{-1.4}$, consistent with the entire observed disk (not just the
simulated inner part) having mass $10^4 M_\odot$.  Each particle
starts with an eccentricity and inclination randomly drawn from
uniform distributions between 0 and 0.3 and between 0 and $10^\circ$,
respectively.  The orbital nodes, arguments of pericenter, and mean
anomalies are drawn from a uniform distribution between 0 and $2\pi$.
In this simulation, 6 test particles (low mass stars) are delivered to
orbits with pericenters less than or equal to that of the infalling
cloud between 4 and 6 Myr.  The massive perturbers are only mildly
excited from their initial disk configuration.  In additional
simulations (not shown) starting with lower eccentricities and
inclinations, the perturbers self-excite to values comparable to those
observed on our timescale of interest.  The number of stars delivered
before 6 Myr onto orbits with pericenter distances less than $r_p =
270$AU varies substantially from simulation to simulation and can
include 0.  To elucidate the orbital evolution of the stars, we
perform a longer simulation, this time with 24 perturbers having $m_*
= 100 M_\odot$ and 940 test particles.  The results are presented in
Supplementary Figure S3.  At early times, the distribution of angular momenta is well
matched by a 1-D diffusion equation, with the number plotted
\begin{eqnarray}\label{eqn-diff}
N\left(\log_{10}\frac{J}{J_c}\right) = (4\pi Dt)^{-1/2} e^{-(J-J_c)^2/(4Dt)} J\ln(10) N_* \Delta \;, ~~~~~~~~~~~~~~~~~~~~~~(S5) \nonumber
\end{eqnarray}
where $\Delta$ is the binsize.  To calculate $D$, we use equation (S3)
with $R = 10^4$AU, $h/R = 10^\circ$, $\zeta =\pi$ and $\ln\Lambda =
7$.  At late times, phase space fills and $N = N_* (J/J_c)^2 \Delta$,
as expected.  At 6 Myr, the transition between these regimes is
marginal, and small number statistics generate a large variation in
outcomes given our fiducial parameters.

The expected closest approach distance, $b_{\rm min}$, and the rate of
orbital evolution, $f$, are related by $f \sim T^{-1} pN_* (b_{\rm
strong}/b_{\rm min})^2 \ln{\Lambda} $, where $p = v_{\rm
rand}^2/(\Omega R)^2 = e_*^2$.  Rearranging, $b_{\rm min} \sim b_{\rm
strong} (pN_*/N_{del})^{1/2}(\ln{\Lambda})^{1/2}$, where $N_{del}
\equiv fT$ is approximately the number of stars delivered onto
plunging orbits over the age of the system, $T$.  For simplicity, we
set $\ln\Lambda = 7$ since variations in $b_{\rm min}$ considered here
will not substantially alter this value.  These choices give $b_{\rm
min} \sim b_{\rm strong}$ if all stars are delivered onto plunging
orbits over the system age $T$ and $b_{\rm min}$ is larger if only a
fraction of the population is delivered.  Disk disruption is avoided
if the delivery fraction $N_{del}/N_* < (1/8)(\ln{\Lambda})(m_{\rm
pert}/m_*)v_{\rm th}^3/(v_{\rm rand}\Omega^2 R^2) = 0.01(m_{\rm
pert}/60 M_\odot)(0.3 M_\odot/m_*)$, where we have chosen to use the
expression for early times since low velocity encounters are more
disruptive.  For reference, in the simulation shown in Supplementary Figure S2, 2/6
of the stars delivered onto plunging orbits at ages between 4 and 6
Myr do not experience disruptive encounters.

Hence, $N_*/300 = 0.5$ solar-mass stars and about $N_*/100 = 10$ stars with $m_* = 0.3 M_\odot$ can be delivered over the disk lifetime T without expecting a disk disrupting encounter.  For $T = 6$ Myr, the rate $f$ above generates delivery of about $N_*/50$ of each stellar population, for a total of 3 solar mass stars and 20 $0.3 M_\odot$ stars.  These numbers are comparable within the errors in our calculation, and disks can be safe from disruption. 

\noindent
{\bf Other scatterers.}  What about disruption from encounters with
lower-mass stars or with high mass perturbers in an isotropic
population?  The rate of encounters generating a fixed impulse $\Delta
v$ to the disk scales as $N_{\rm pert}m_{\rm pert}^2/(\sigma_\star
v_{\rm rel})$ for a fixed stellar mass.  In a given dynamical
population, high mass stars dominate disk disruption, so we need not
worry about encounters with low-mass stars in the young disk.  For an
isotropic population, $(\sigma_\star v_{\rm rel})^{-1}$ is
approximately 0.13 times that for the disk population, so both the
star's orbital evolution rate $f$ and the disk disruption rate scale
as 0.13$N_{\rm pert} m_{\rm pert}^2$.  Dynamical limits constrain the
mass of stars and stellar remnants in the galaxy's central $0.1$ pc to
be $\lesssim$$10^5 M_\odot$.\cite{f15} The expected number of low-mass
stars and stellar to intermediate mass black holes in this region is
not clear.\cite{f18} The diffusion of lighter stars out of the ring
may have led to the unusually top-heavy mass function of stars that
are found in the ring at
present,\cite{f15,f27,f32} and more generally, the
isotropic distribution likely contains the remnants of older star
formation episodes.  For example, $10^5$ $1 M_\odot$ perturbers in an
isotropic population would disrupt disks at $\sim$1/10 the rate of our
fiducial 60$M_\odot$ perturbers.  A large population of
intermediate-mass black holes, in contrast, could generate higher disk
disruption rates, but the number of such objects, if they exist, is
unconstrained and need not be high enough to cause problems for our
scenario.

Finally, we return to the probability of a final strong encounter that
altered the plunging star's pericenter distance from $2r_{\rm p}$ to
$r_{\rm p}$.  The ratio of scattering events at $b_{\rm min}$ to weak
scatterings is $1/(\ln\Lambda)$, so about 1/7 injected stars will come
in with a strong (i.e., $b \sim b_{\rm min}$) final scattering
event. The rate of orbital evolution due to ``ideal" strong
scatterings with a favorable encounter velocity due to the encounter
geometry is increased relative to the rate of typical strong
scatterings by the ratio of the typical to the favorable encounter
velocity (gravitational focusing triumphs) and is decreased by the
fraction of time spent by a perturber with a favorable velocity.  How
likely strong scatterings are to occur at favorable encounter
velocities depends on the details of the distribution of perturbers,
but since we prefer encounters near perturbers' apocenters, where they
spend the most time, favorable encounter geometries are not rare.

If one IMBH having mass $M_{\rm IMBH} = 1.5\times10^3M_\odot$ resides
in the central R = 0.04 pc of the galaxy, then the likelihood of a
strong encounter at $b_{\rm strong} = GM_{\rm IMBH}/\sigma_\star^2 =
3$AU with an isotropic $\sigma_\star \approx 670~{\rm km~s^{-1}}$,
capable of sending a star plunging directly from a circular orbit, is
approximately $1/( 4\pi R^3/3)\pi b_{\rm strong}^2 \sigma_\star T =
0.01$ per star for $T = 6\times 10^6$ years.  This corresponds to 9
injections over the disk lifetime for $N_* = 900$.  Stellar encounters
with an IMBH at 3AU generate substantial disk stripping.  The
probability per star of a maximal non-stripping encounter with the
IMBH approaches 1, and each such encounter can generate $\Delta v_* \sim
150 (0.3 M_\odot/m_*)^{1/2}~{\rm km~s^{-1}}$.  One or more IMBHs in
the central region could substantially increase the rate of diffusion
of low mass stars onto low angular momentum orbits.

\noindent
{\bf Summary.} In conclusion, approximately 1/100 of the population of
low-mass stars in the young stellar ring could have been delivered by
the observed population of 60 $M_\odot$ perturbers onto orbits
comparable to that of the observed plunging cloud over the age of the
young ring without disk stripping.  A few thousand young low-mass
stars could be present in the ring, consistent with limits from X-ray
observations,\cite{f30} of which about a quarter likely lie in the
inner ring.  These constraints allow approximately $2\times10^{-6}$
injections of undisrupted disks per year.  We conclude that a tail
population of low-mass (previously unobserved) stars could diffuse
into orbits similar to that of the gas cloud around SgrA*.  Is this
delivery rate sufficient?  A disk orbiting a 0.3$M_\odot$ star must
have a pericenter distance approximately 2--3 times larger than that
of the current cloud in order to remain undisrupted at $d_{\rm esc}$.
On such an orbit, the disk will remain bound indefinitely, and the
wind will be regenerated over each orbital period, making the
probability of observing such an object in the Galactic centre close
to unity.  Our calculations are sufficiently approximate that the
infalling cloud may represent a long-lived, tidally truncated disk
with a continually regenerated wind.  If, on the other hand, we need a
final kick that recently reduced the pericenter distance of the cloud
by a factor of two, that kick needed to generate $\Delta J/J = 1/4$,
as calculated above.  Such final strong kicks will occur at
approximately 16/7 of the delivery rate, where the factor of 16 comes
from $(J/\Delta J)^2$ and the factor of 7 comes from the Coulomb log.
Over the 140 year period of the cloud, the probability that an
infalling disk reaches the current cloud pericenter for the first time
is $\sim$0.1\%.  If the cloud is indeed on its first plunge after a
pericenter modification of a factor of two, we are somewhat fortunate
to observe it.

Our calculated probability is high enough that observing a plunging
young star hosting a proto-planetary disk is plausible.  We do not
argue that the presence of such an object is likely {\it a priori},
but rather than the observed young cloud has properties that can be
naturally explained given our scenario, and furthermore, our model
generates predictions about the future evolution of the cloud that are
not common to the pressure-confined cloud model proposed by the
discovery team.\cite{f1} If our scenario is correct, photoevaporating
proto-planetary disks with pericenter distances 2--3 times that of the
infalling cloud should be common.

We note that if our suggestion is correct and observations over the
next few years confirm that the infalling gas cloud arises from the
tidal evolution of a wind blowing off a circumstellar disk, more
exotic sources for such a disk need not be considered.  For example, a
grazing collision between a star and a stellar mass black hole could
both put the star onto a sufficiently eccentric orbit and generate
debris, a small fraction of which might settle into a disk around the
star.\cite{f33} However, we consider such a history less natural than
our interpretation of the infalling cloud as a young proto-planetary
disk.

\noindent
{\bf Other similar systems.} Several previous observations in the
region of the Galactic centre might also be attributable to
proto-planetary disks.  Clenet et al.~found a flaring infrared source
in the $L'$ band at a projected separation of $\sim$0.003pc = 670 AU
from SgrA*.\cite{f34} The emission is not likely to be stellar since
no $K$-band counterpart was detected (implying $(K-L^\prime)>3$) and
the source appears extended. If the emission is associated with dust,
it corresponds to a dust temperature of $\sim 10^3$K with a total
luminosity of $\sim$$10L_\odot$. It is possible that this source is of
similar origin to the observed infalling cloud considered here, for
which $K_s$, $L^\prime$, and $M$ band fluxes imply emission
from\cite{f1} dust with a temperature of $550 \pm 90$K and a total
luminosity of $\sim$$5L_\odot$.

Fritz et al.~found that 9 out of 15 objects detected in the $K_s$ and $H$-bands in the star complex IRS 13E ($\sim$0.14pc $=  3\times 10^4$ AU from SgrA*, with a core radius of about 2500AU) are very red, consistent with
them being warm dust clumps.\cite{f35} In principle, these dust clumps could be
formed by the evaporation of proto-planetary disks around young stars in
this complex.

Finally, Muzic et al.~reported another star complex IRS 13N at a similar projected separation from SgrA*.\cite{f36}
This complex has a radius of $\sim$2000 AU and includes extremely red sources with colors of either
dust-embedded stars older than a few Myr or extremely young stars with
ages $\lesssim 1$ Myr. The latter interpretation is supported by the fact
that six of the sources are close in projection and show very similar
proper motion whose coherence is not expected to survive over an orbital
time. Such young stellar objects could naturally host proto-planetary
disks of the type required to explain the infalling cloud considered here.

Compact star clusters such as IRS 13N (or IRS 13E), with mass
estimates of a few thousand solar masses, could also serve as massive
perturbers able to scatter a low-mass star onto the orbit of the
plunging cloud.

\small
\noindent
\begin{suppbibliography}{}

\bibitem{f23} Jackson, J.M. \textit{et al.}, Neutral
Gas in the Central 2 Parsecs of the Galaxy, {\it Astrophys. J.} {\bf
402}, 173-184 (1993).

\bibitem{f24} Merritt, D., Gualandris, A. \&
Mikkola, S., Explaining the Orbits of the Galactic Center S-Stars,
{\it Astrophys. J.} {\bf 693}, L35-L38 (2009).

\bibitem{f25} Merritt, D., \& Wang, J., Loss Cone
Refilling Rates in Galactic Nuclei, {\it Astrophys. J.} {\bf 621},
L101-L104 (2005).

\bibitem{f26} Wang, J. \& Merritt, D., Revised Rates of Stellar Disruption in Galactic Nuclei, {\it Astrophys. J.} {\bf 600}, 149-161 (2004).

\bibitem{f27} Bartko, H. \textit{et al.}, An
Extremely Top-Heavy Initial Mass Function in the Galactic Center
Stellar Disks, {\it Astrophys. J.} {\bf 708}, 834-840 (2010).

\bibitem{f28} Bastian, N., Covey, K.R., \& Meyer,
M.R., A Universal Stellar Initial Mass Function? A Critical Look at
Variations, {\it Annu. Rev. Astro. Astrophys.} {\bf 48}, 339-389
(2010).

\bibitem{f29} Bonnell, I. A., \& Rice, W. K. M.,
Star Formation Around Supermassive Black Holes, {\it Science} {\bf
321}, 1060 (2010).

\bibitem{f30} Nayakshin, S. \& Sunyaev, R., The
`missing' young stellar objects in the central parsec of the Galaxy:
evidence for star formation in a massive accretion disc and a top
heavy initial mass function, {\it Mon. Not. R. Astron. Soc.} {\bf
364}, L23-L27 (2005).

\bibitem{f31} Chambers, J.E., A hybrid symplectic
integrator that permits close encounters between massive bodies, {\it
Mon. Not. R. Astron. Soc.} {\bf 304}, 793-799 (1999).

\bibitem{f32} Manness, H. \textit{et al.}, Evidence
for a Long-standing Top-heavy Initial Mass Function in the Central
Parsec of the Galaxy, {\it Astrophys. J.} {\bf 669}, 1024-1041 (2007).

\bibitem{f33} Miralda-Escud\'e, J., A star disrupted
by a stellar black hole as the origin of the cloud falling towards the
Galactic center. Preprint at http:// arXiv.org/abs/1202.5496 (2012).

\bibitem{f34} Clenet, Y. \textit{et al.}, A dual
emission mechanism in Sgr A*/L'?, Astron. \& Astrophys. {\bf 439},
L9-L13 (2005).

\bibitem{f35} Fritz, T. K. \textit{et al.},
GC-IRS13E---A Puzzling Association of Three Early-Type Stars,{\it
Astrophys. J.} {\bf 721}, 395-411 (2010).

\bibitem{f36} Muzic, K. \textit{et al.}, IRS 13N: a
new comoving group of sources at the Galactic center, Astron. \&
Astrophys. {\bf 482}, 173-178 (2008).

\end{suppbibliography}

\end{document}